\documentclass[twocolumn,prd,floatfix,nofootinbib,longbibliography,notitlepage,superscriptaddress]{revtex4-2}

\pdfoutput=1

\usepackage{amsmath, amssymb, amsfonts, amsthm, latexsym, mathrsfs, bbm, slashed}

\usepackage[usenames,dvipsnames]{xcolor}
\usepackage[title,titletoc]{appendix}
\usepackage{graphicx}
\usepackage{xifthen}
\usepackage{tensor}
\usepackage{dsfont}
\usepackage{physics}
\usepackage[inline]{enumitem}

\usepackage{setspace}
\usepackage[marginal, multiple]{footmisc}
 
\usepackage[T1]{fontenc}
\usepackage[utf8]{inputenc}
\usepackage{lmodern}
\usepackage[normalem]{ulem}

\usepackage[unicode, colorlinks, allcolors=blue!70!black, linktocpage, pdfusetitle]{hyperref}
\definecolor{CiteColor}{rgb}{0.55,0,0}
\hypersetup{citecolor=CiteColor}
\definecolor{RefColor}{rgb}{0,0.5,0}
\hypersetup{linkcolor=RefColor}
\usepackage[all]{hypcap}

\usepackage{orcidlink}
\usepackage{braket}

\usepackage{verbatim}

\usepackage{cancel}

\usepackage{microtype}
\usepackage{standalone}
\usepackage{tikz}
\usetikzlibrary{arrows.meta}
\usetikzlibrary{decorations.markings}
\ifdefined\myext
\usetikzlibrary{external}
\fi

\newcommand{\pd}{\partial}

\mathchardef\minus = "002D

\newcommand{\scA}[4][]{{}_{{}_{#2}}A^{#1}_{#3}(#4)}

\newcommand{\cf}{\mathcal{C}}

\newcommand{\Aa}{\scA{s}{\ell{m}}{c}}
\newcommand{\A}{{ }_{s} A_{\ell m}}

\begin{document}

\title{Towards a more robust algorithm for computing the Kerr quasinormal mode frequencies}

\newcommand{\UMiss}{\affiliation{Department of Physics and Astronomy, The University of Mississippi, University, MS 38677, USA}}

\author{Sashwat Tanay\,\orcidlink{0000-0002-2964-7102}}
\email{stanay@go.olemiss.edu}
\UMiss
%\author{Leo C.~Stein\,\orcidlink{0000-0001-7559-9597}}
%\email{lcstein@olemiss.edu}
%\UMiss

% Because hyperref only gets the *last* author, we need to be explicit.
\hypersetup{pdfauthor={Tanay, Cho, and Stein}}

\begin{abstract}
Leaver's method has been the standard for computing the quasinormal mode (QNM) frequencies for a
Kerr black hole (BH) for a few decades. We start with 
a spectral variant of Leaver's method introduced by
Cook and Zalutskiy~[\prd~90, 124021 (2014)],
 and propose improvements in the form of computing the 
necessary derivatives analytically, rather than by numerical finite differencing.
We also incorporate this derivative information into \texttt{qnm}, a \textsc{Python}
package which finds the QNM frequencies via the spectral variant of Leaver's method.
We confine ourselves to first derivatives only.
\end{abstract}

\maketitle
%% no need for toc; small paper
%\tableofcontents

\section{Introduction}    \label{intro}

When the two component black holes (BHs) of a binary black hole
(BBH) system merge together to form a single BH, this BH
oscillates in the so-called quasinormal modes (QNMs).
This final state of the system is referred to as the ringdown state,
 in contrast to the initial inspiral state 
wherein the two component BHs of the system slowly rotate around a common center.
All this while, the BBH system and and resulting single BH keep on emitting 
gravitational waves (GWs). The construction of templates for accurate parameter estimation
using GWs requires the modeling on the ringdown state as well.
The QNM frequencies of 
a Kerr BH are functions of its parameters like mass and spin.
All this makes the determination of the QNM frequencies becomes a matter of 
importance.

The QNMs have a fifty-plus years long history; see 
Refs.~\cite{Berti:2009kk, Nollert_1999} for reviews on the 
QNM literature. We will confine ourselves to the 
literature immediately useful for this paper. 
Leaver's method has become the standard 
way to hunt for the Kerr QNM frequencies \cite{Leaver:1985ax},
although modifications have been suggested over the years \cite{CZ}.
The method starts with Teukolsky equations which describe the perturbations 
to the Kerr geometry. One then feeds a Frobenius series ansatz  
(representing these perturbations) to the Teukolsky equations,
 both in the radial
and angular sectors.
This leads to recurrence relations which then are turned into two infinite
continued fraction (CF) equations.
Complex $\omega$ (the QNM frequency) and a separation constant $A$
are the two roots of these two CF equations which are to be numerically found.
All this is done for a fixed dimensionless
BH spin parameter $a$, where $0 < a < 1$.

Cook and Zalutskiy in Ref.~\cite{CZ}, presented a variant of Leaver's method where 
the CF approach was retained while dealing with the radial sector but 
spectral decomposition was introduced in the angular sector, 
thereby recasting the problem as 
a CF equation and an eigenvalue equation, both coupled together;
also see Appendix A of Ref.~\cite{Hughes:1999bq}.
The advantage of the spectral approach is that it 
leads to rapid convergence towards the solution.
%Recently, one of us (L.C.S) released \texttt{qnm}, a 
%\textsc{Python} package with Ref.~\cite{Stein:2019mop}
% implementing the spectral approach of Ref.~\cite{CZ}.
 Recently, \texttt{qnm}, a  \textsc{Python} package
 was released with Ref.~\cite{Stein:2019mop} which
 implemented the spectral approach of Ref.~\cite{CZ}.

This paper aims to establish some theoretical foundations on which improvements to the 
spectral variant of Leaver's method of Ref.~\cite{CZ} can be brought about. 
This is mainly done by introducing ways
 to replace the computation of numerical derivatives 
(via finite differencing) with those of analytical ones.
 This is so because in general, 
analytical derivatives are more accurate and
 reliable than their numerical counterparts.
In this paper, we confine ourselves to computing only the first derivatives,
while leaving second derivatives for future work.
We also incorporate these improvements in the \textsc{Python}
package \texttt{qnm}.

The organization of the paper is as follows. 
The problem of finding the QNM frequencies of a Kerr BH is introduced in 
Sec.~\ref{setup} as a root-finding and an eigenvalue problem, coupled 
together. In Sec.~\ref{standard_sec}, we highlight some aspects of the present-day
methods of computing these frequencies where the inclusion of 
analytical derivative information could be useful.
Then in Sec.~\ref{optimize}, we show how to compute the necessary derivatives analytically,
while leaving some  calculational details to Sec.~\ref{calc_details}.
We devote some time to discussing the implementation of the derivative information
in the \textsc{Python} package \texttt{qnm} in Sec.~\ref{python_package}, before
summarizing in Sec.~\ref{conc}.

\section{QNMs  as roots of a continued fraction equation}
\label{setup}

We will try to isolate the reader from some general relativistic and 
other mathematical considerations while casting the problem of
finding the QNM frequencies of a Kerr BH as a root-finding problem
and an eigenvalue problem, coupled together.
The interested reader is referred to 
Ref.~\cite{CZ} and the references therein for details.
Our starting points will be Eqs.~(44) and (56) of Ref.~\cite{CZ},
which we reproduce here (with $c \equiv 	a \omega$, where
 $a$ is the dimensionless BH
spin parameter and $\omega$ is the sought-after QNM frequency).
\begin{align}            \label{cf_eqn}
& {\cf}(\omega, \A, a, s; {n},  {N})      \nonumber   \\
& \equiv  \beta_{n}-\frac{\alpha_{n-1} \gamma_{n}}{\beta_{n-1}-} \frac{\alpha_{n-2} \gamma_{n-1}}{\beta_{n-2}-} \ldots \frac{\alpha_{0} \gamma_{1}}{\beta_{0}}  \nonumber  \\
&-\frac{\alpha_{n} \gamma_{n+1}}{\beta_{n+1}-} \frac{\alpha_{n+1} \gamma_{n+2}}{\beta_{n+2}-} \ldots \frac{\alpha_{N-1} \gamma_{N}}{\beta_{N}+\alpha_{N} r_{N}}   = 0  
\end{align}
\begin{align}
\mathbb{M} \cdot \vec{C}_{\ell m}(c)={ }_{s} A_{\ell m}(c) \vec{C}_{\ell m}(c)     \label{eigval_eqn}
\end{align}
A few clarifying comments are warranted at this point. First of all,
unlike Ref.~\cite{CZ}, we will use $\cf$ instead of
 $\text{Cf}$ to denote the CF
or any of its inversions.
${\cf}(a, \A, \omega, s; {n} , {N}) $ denotes the $n$th inversion of the truncated CF
\begin{align}
&  \cf(a, \A, \omega, s;  {N})    \nonumber   \\   
& \equiv   \cf(a, \A, \omega, s; 0,  {N})        \nonumber   \\
   &  \equiv  \beta_{0}-\frac{\alpha_{0} \gamma_{1}}{\beta_{1}-} \frac{\alpha_{1} \gamma_{2}}{\beta_{2}-} \frac{\alpha_{2} \gamma_{3}}{\beta_{3}-} \ldots \frac{\alpha_{{N}-1} \gamma_{{N}}}{\beta_{{N}}+\alpha_{{N}} {r}_{{N}}}.
\end{align} 
Next, $\alpha_n$'s, $\beta_n$'s and $\gamma_n$'s 
 are functions of  $a$, 
$\Aa$ (the eigenvalue in Eq.~\eqref{eigval_eqn}), $\omega$,
and spin of the gravitational field $s$.
For fixed values of $m, s$ and $c \equiv a \omega$, we have the matrix
$\mathbb{M}$ whose eigenvalues $\Aa$'s are labeled by $\ell$.
Altogether, this means that for fixed values
of $s, \ell, m$ (discrete parameters) and $a$ (continuous parameter), Eqs.~\eqref{cf_eqn}
and \eqref{eigval_eqn} are two equations to be solved for the eigenvalue
$\A$ and root $\omega$.
This point of view gives meaning to expressions like
$\omega(a)$ or $\A(a)$.
Stated differently, for fixed values
of $s, \ell, m$ and $a$,  Eq.~\eqref{cf_eqn} is a root-finding
problem in $\omega$ (if $\A$ is fixed), 
and Eq.~\eqref{eigval_eqn} is an eigenvalue problem for the eigenvalue
$\A$ (if $\omega$ is fixed).

For fixed and physically relevant values of $(s, \ell, m, a, \omega)$, 
(which implies a fixed $\A$),
there are an infinite number of QNM frequencies $\omega$'s (along with the $\A$) that 
satisfy Eq.~\eqref{cf_eqn}. 
These $\omega$'s are generally labeled with another (overtone) index 
$p = 0, 1, 2, 3 \cdots$.
In general, the $p$th overtone QNM frequency  $\omega$ is the most 
stable root of the $p$th inversion of the CF $\cf(p,N)$ .
We will sometimes denote $\A$ as simply $A$, thus suppressing
the dependence on $s, \ell$ and $m$. This concludes our brief 
review of casting the problem of finding QNM frequencies as that
of a coupled root-finding problem (the root being $\omega$) and an eigenvalue problem
(the eigenvalue being $A$), involving 
Eqs.~\eqref{cf_eqn} and \eqref{eigval_eqn}.

\section{Spectral variant of Leaver's method}      \label{standard_sec}

We now briefly highlight some aspects of computations of QNM frequencies adopted in 
 Refs.~\cite{CZ, Stein:2019mop} which require computation of certain derivatives.
In the context of Eq.~\eqref{eigval_eqn}, we may write $A(c)$ as $A(a, \omega)$
since $c \equiv a \omega$.
With this, we can display the functional dependencies involved in Eq.~\eqref{cf_eqn} 
as\footnote{We have suppressed the display of $(n, N)$ 
dependence of $\cf$ in the above equation because it is irrelevant to the
ongoing discussion.}
\begin{align}
\mathcal{C} (\omega ,~ A(c(a, \omega  ))~ , a) = \mathcal{C}(a, \omega)   = 0   .      \label{Newt-Raph-eqn}
\end{align}
The main point is that Eq.~\eqref{Newt-Raph-eqn} expresses
a parameterized (by $a$) root-finding problem
in a single complex variable $\omega$.

A few key aspects of the standard approach to computing QNM frequencies
are as follows
\begin{enumerate}
\item  The derivative $\pd \cf/ \pd \omega$ needed to find 
the slope of $\cf (a, \omega)$ for the Newton-Raphson procedure, is computed numerically via finite differencing.
% We explain in Appendix \ref{diff_CF} why computing this derivative analytically is not straightforward.
\item After having found the solutions $\omega_0(a_0)$ and $A_0(a_0)$ at BH spin $a = a_0$, we move to the next
value of the BH spin $a = a_0 + \Delta a_0$. 
The initial guesses  $\omega_{\text{guess}}$ and $A_{\text{guess}}$
for  $a = a_0 +  \Delta a_0$ is usually constructed by a quadratic extrapolation using  solutions 
at three previous values of $a$. Note that this quadratic extrapolation is the numerical finite-differencing
equivalent
of taking the second derivatives of $\omega(a)$ and $A(a)$ with respect to $a$.
\end{enumerate}
The point of this paper is to suggest analytical replacements to the above-mentioned
instances where finite-differencing is employed to compute the derivatives in the standard approaches. This is so because
finite-differencing is considered as an unreliable way to compute the 
derivatives\footnote{The reader is referred to Secs.~5.7 and 9.4 of Ref.~\cite{press2007numerical}
for a detailed discussion on why
finite-differencing is not the preferred way to compute the derivatives 
in general and
particularly so in the context of the Newton-Raphson method, respectively.}.

\begin{figure}[t]
  \includegraphics[width=9.5cm]{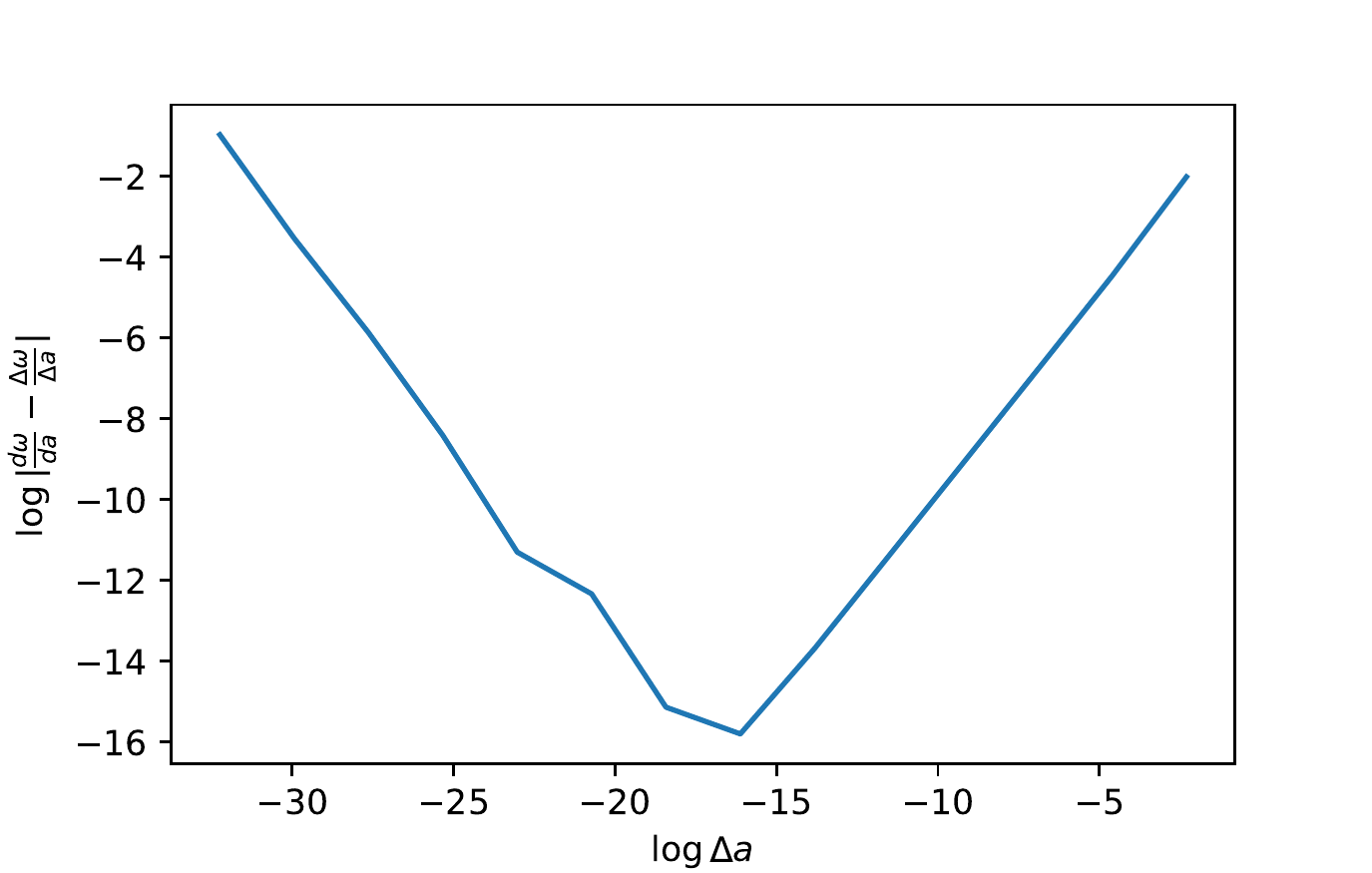}
  \caption{Natural logarithm of the modulus of the difference between the analytical and the finite-difference
versions of $d \omega/da_{\text{analytical}}$ as a function of the
natural  logarithm of the step size. This is done for 
$a = 0.68, s = -2, m = 2$ and the overtone index 
$ p = 5$. 
} 
  \label{fig:1} 
\end{figure}

There is an interesting aspect about supplying the initial guesses for $(\omega, A)$ to the
root-finding routine that warrants some discussion.
If we use the previous solution for $(\omega_0, A_0)$ at $a =a_0$ as a guess for the solution at 
$a = a_0 + \Delta a_0$, then we found from numerical inspection with a 
handful of cases that the maximum
step size in $a$ that we can take is $ \Delta a_0 \sim 0.02$.
 A larger step-size results in the 
numerical procedure not being able to find the root or ending up on a different root than what it is
supposed to. When the first derivative information 
($d \omega/da$ and $dA/da$) is supplied
to determine $\omega_{\text{guess}}$ and 
$A_{\text{guess}}$ for the next value of $a$ 
(via linear extrapolation), we found that
we can safely take a step size of $ \Delta a_0 \sim 0.25$.
 When the (numerical) second derivative information
  is supplied (via quadratic extrapolation),
a step size of even $ \Delta a_0 \sim 0.65$ looks possible.
 The above findings indicate that including the  derivative information
to generate $\omega_{\text{guess}}$ and $A_{\text{guess}}$ at the next value of $a$
can allow us to take much larger steps 
$\Delta a$ than would be possible otherwise.
One should note that although the derivative information 
 lets us take large step size in $a$, but accurate numerical
computation of these derivatives requires us to take reasonably small step size 
to begin with so that the truncation error is reduced.
This is against the spirit of being able to take large steps $\Delta a$. 
Also when the step size gets too small, roundoff error starts to increase.
All this is clearly displayed in Fig.~\ref{fig:1}, where we plot
the natural logarithm of the modulus of the difference between the analytical
(via Eq.~\ref{central_eqn}) and the finite-difference
versions of $d \omega/da_{\text{analytical}}$ as a function of the natural logarithm
of the step size. This difference initially decreases with a decrease in the step size
(decreasing truncation error) $\Delta a$, but then starts to increase (increased roundoff errors).
See the caption of Fig.~\ref{fig:1} for more details\footnote{Fig.~\ref{fig:1} has been plotted
 with 16 digits of machine precision. The root-finding routine had an absolute tolerance of 
 $\sim 10^{-8}$ on the function whose zero is to be found. To compute the CFs for this figure,
 the CF-computing routine had an absolute tolerance of $\sim 10^{-10}$ to check the convergence
 of the CF.}.
All this demonstration suggests that having the derivatives analytically 
is certainly desirable.

\section{Improvements to the root-finding procedure}     \label{optimize}

\subsection{General considerations}

For a certain $a$, Eq.~\eqref{Newt-Raph-eqn} when solved for the QNM frequency via numerical 
root-finding,  gives a certain $\omega$\footnote{There are an infinity of $\omega$'s that satisfy Eq.~\eqref{Newt-Raph-eqn}
for a certain $a$.}. It is in this sense that we can write $\omega$ as a function 
of $a$; i.e. $\omega(a)$. 
This further allows us to write  Eq.~\eqref{cf_eqn} as
\begin{align}
\cf(\omega, A, a) =  \mathcal{C} (\omega(a),~ A(c(a, \omega(a)))~ , a) = \mathcal{C}(a)    = 0,   \label{cf_eqn_1}
\end{align}
Since  Eq.~\eqref{cf_eqn_1} views $\mathcal{C}$ as
 a function of just one variable $a$, so that $d \mathcal{C}/da$
 is a sensible quantity. Thus we have
 \begin{align}
 \frac{d \cf}{da}   = &  \frac{\pd \cf}{\pd \omega}  \frac{d \omega}{da}  +
 \frac{\pd \cf}{\pd A}  \frac{d A}{da} + \frac{\pd \cf}{\pd a}  = 0    \\ 
  & \frac{\pd \cf}{\pd \omega}  \frac{d \omega}{da}  +
 \frac{\pd \cf}{\pd A}  \left[   \frac{dA}{dc} \left(  \frac{\pd c}{\pd a} + \frac{\pd c}{\pd \omega} \frac{d \omega}{da} \right)    \right]+ \frac{\pd \cf}{\pd a}  = 0    \\ 
  & \textcolor{black}{\frac{\pd \cf}{\pd \omega} } 
  \textcolor{black}{\frac{d \omega}{da}} +
\textcolor{black}{ \frac{\pd \cf}{\pd A} } \left[   \frac{dA}{dc} \left(  \omega+a   \textcolor{black}{\frac{d \omega}{da}} \right)    \right]+ \textcolor{black}{\frac{\pd \cf}{\pd a}}  = 0  
\label{pre_central_eqn}   \\
\frac{d \omega}{da}   =  &    \frac{ - \pd \cf/\pd a - \omega ~ dA/dc  ~ \pd \cf/\pd a  }{a ~ dA/dc ~ \pd \cf /\pd A  + \pd \cf/ \pd \omega}  . \label{central_eqn}
 \end{align}
We also reproduce the expression of  $dA/da$ in square braces in Eq.~\eqref{pre_central_eqn} for
this, in addition to Eq.~\eqref{central_eqn} will be useful
for providing $\omega_{\text{guess}}$ and $A_{\text{guess}}$ at the next spin $a +
\Delta a$ via linear extrapolation
\begin{align}
\frac{dA}{da}  =  \frac{dA}{dc} \left(    \omega   +  a \frac{d \omega}{da} \right).    \label{central_eqn_2}
\end{align}

\subsection{Improvements to the pre-existing algorithm of finding QNMs}   \label{our-method}

The first improvement we propose is to provide the Newton-Raphson routine
 the $\omega$-derivative of the CF $\cf$ in a closed-form.
The $\omega$-derivative of $\cf$ in Eq.~\eqref{Newt-Raph-eqn} is computed as
(with $a$ dependence suppressed because the context is that of a fixed BH spin $a$)
\begin{align}
\label{eq:Newton-derivative}
  \frac{d \cf(\omega, A(\omega))}{d \omega}   &  =   \frac{\partial  \cf ( \omega,A)}{\partial \omega}
  + \frac{\partial \cf (\omega,A)}{\partial A} \frac{d A}{d \omega}  ,        \\
  &  =   \frac{\partial  \cf ( \omega,A)}{\partial \omega}
  + a  \frac{\partial \cf (\omega,A)}{\partial A} \frac{d A}{d c}     . \label{dCdomega_5}   
\end{align}

The second proposal is in regard to the starting guesses
(in the context of numerical root-finding) for 
QNM frequency $\omega$ and the eigenvalue $A$.
 We propose that at $a_0 + \Delta a$, the guesses for 
 $\omega$ and $A$ should be
 \begin{align}
 \omega_{\text{guess}}  = \omega_0 + \frac{d \omega}{da} \bigg\vert_{a_0, \omega_0}  da   ,     \label{omega_guess}  \\
 A_{\text{guess}}  = A_0 + \frac{d A}{da} \bigg\vert_{a_0, A_0}  da  ,  \label{a_guess}
 \end{align}
where the total derivatives in the above two equations are to be computed using
Eqs.~\eqref{central_eqn} and \eqref{central_eqn_2}.
Now, to actually implement the above two improvements suggested in Eqs.~\eqref{dCdomega_5}, \eqref{omega_guess}, and \eqref{a_guess},
we need to compute $dA/dc$ and the
three partial derivatives of $\cf$, i.e., $\pd \cf/\pd a, \pd \cf/\pd A$ and $\pd \cf/\pd \omega$.
This is the subject of the next section.

In light of Eqs.~\eqref{Newt-Raph-eqn} and \eqref{dCdomega_5}, 
we can discuss one more application of 
analytical derivatives . The QNM frequency is found by finding the root of
$\cf (\omega) = 0$; here the $a$-dependence of $\cf$ is suppressed.
Now the software routines (like \texttt{qnm} \cite{Stein:2019mop}), have 
separate tolerances for checking the convergence of the CF $\cf$, and numerical root-finding 
for $\cf (\omega) = 0$, and are 
usually set independently of each other. But now with the 
$d \cf/d\omega$ information from Eq.~\eqref{dCdomega_5},
 we can  make the former depend on the latter.
This can be done by setting
\begin{align}
\Delta \cf  \sim  \Delta \omega   \frac{d \cf}{d \omega},
\end{align}
where the symbol $\Delta$ before $\cf$ and $\omega$ represents their respective tolerances.
 $\Delta \omega$ is usually decided on the basis of how accurately we want to determine 
 the QNM frequencies, something that we need to consider while doing
 precise parameter estimation with GWs.

\section{Calculational details}     \label{calc_details}

\subsection{Derivatives of a continued fraction} \label{diff_CF}

Before we deal with computing
 the derivatives of a CF,
we will first discuss the challenges encountered 
when trying to compute them.
We start with a typical CF 
\begin{equation}
f(x)=b_{0}+\frac{a_{1}}{b_{1}+} \frac{a_{2}}{b_{2}+} \frac{a_{3}}{b_{3}+} \frac{a_{4}}{b_{4}+}  \cdots ,
\end{equation}
where the quantities $a_n$'s and $b_n$'s are considered to be functions
of a parameter $x$, and the aim is to differentiate $f(x)$ with respect 
to $x$. It is not hard to see that straightforward  application 
of the basic rules of differentiation
(product, quotient, and chain rules) to a finite-truncated version 
of the above CF will result in an explosion of
the size of expression of the derivative, if a large number of terms
 are retained in $f(x)$. The requirement to
store and process this large data on a computer algebra system 
like \textsc{Mathematica} can make the procedure inefficient 
and slow.

There is a  more efficient way to compute the derivative 
of a CF. This rests on Lentz's method
of computing CFs (see Sec.~5.2 of Ref.~\cite{press2007numerical}
or Ref.~\cite{aip_lentz}). The
 method dictates the following procedure to compute the
 CF
\begin{enumerate}
\item Set $f_0 (x) = b_0(x)$, $C_0(x) = f_0(x)$ and $D_0(x) = 0$. 
\item For $j = 1, 2, \cdots$, we perform the following iterative procedure
to obtain $f(x)$ to higher and higher accuracy (denoted by $f_j (x)$) \\
\begin{align}      
D_{j}(x) & = 1 /\left(b_{j}(x)+a_{j}(x) D_{j-1}(x)\right),    \label{Lentz1}  \\   
C_{j}(x)   & =   b_{j}(x)+a_{j}(x) / C_{j-1}(x),      \label{Lentz2}  \\ 
 f_{j}(x) & =f_{j-1}(x) C_{j}(x) D_{j}(x)         \label{Lentz3}
\end{align}
\end{enumerate}
We terminate when $f_j$ appears to have converged sufficiently well.
The task of processing through each value of $j$ can be called an iteration, which
makes Lentz's method an iterative procedure.
This iterative procedure is the key to its efficiency. Note that
while implementing the $n$th iteration, we need to store
 the information ($a_j$, $b_j$, $C_j$, $D_j$ and $f_j$)
  associated with  $j= n$ and $j = n-1$.
The information corresponding to smaller values of $j$ 
(which were used in the previous iterations)
need not be stored. Apart from computing a CF, this
 efficient iterative procedure  also lets us 
compute the derivative of the CF in a similar 
``Lentz-like'' way, as we now explain.

To compute the derivative of a CF, proceed along the same enumerated steps of 
Lentz's method as above, but after the second step, add a third step to the 
iteration given by
%\begin{widetext}
\begin{enumerate}
\setcounter{enumi}{2}
\item For $j = 1, 2, \cdots$, we perform  \\
\begin{align}      
D'_{j}(x) & =     - \frac{D_{j-1}(x) a'_j(x) + b'_j(x) + a_j(x) D'_{j-1}(x)  }{ ( b_j(x) + a_j(x)  D_{j-1}(x)  )^2 }     ,     \\   
C'_{j}(x)   & =      \frac{1 }{ (C_{j-1}(x))^2}    \left[  C_{j-1}(x) a'_j(x)  \right. \nonumber \\
 & \left.  +  (C_{j-1}(x))^2  b_j'(x) - a_j(x) C'_{j-1}(x)  \right]     ,        \\ 
 f'_{j}(x) & =   C_j(x) D_j'(x) f_{j-1}(x)   \nonumber \\
   & +  D_j(x) (f_{j-1}(x) C_j'(x) + C_j(x) f_{j-1}'(x) ) .      
\end{align}
\end{enumerate} 
%\end{widetext}
The above three equations are results of the application of
basic rules of differentiation on Eqs.~\eqref{Lentz1}-\eqref{Lentz3}.
Here we have assumed that the evaluation 
of $a_j'(x)$ and  $b_j'(x)$ is
tractable. This basically turns the evaluation of $f_j'(x)$ into
an iterative Lentz-like procedure
which we terminate when a reasonable degree of convergence has 
been achieved. This is how we compute 
$\pd \cf/\pd a, \pd \cf/\pd A$ and $\pd \cf/\pd \omega$.

\subsection{Eigenvalue of a perturbed complex-symmetric matrix} \label{QM}

From Eq.~\eqref{eigval_eqn} (an eigenvalue equation),
 we can see that $\A \equiv A$ is a 
$c$-dependent eigenvalue, which further means that 
the task of computing $dA/dc$ is that of computing 
the rate of change of the eigenvalue
with respect to a variable that parameterizes the eigenvalue problem.
This problem appears to be similar to that of quantum mechanical perturbation theory,
or the Rayleigh-Schrodinger perturbation theory (RSPT);
see Chapter 6 of Ref.~\cite{griffiths}.
But there is one important difference. 
In the RSPT, the operator or the matrix
(perturbed and unperturbed) whose eigenvalues are the subject of discussion
is a Hermitian one, whereas here the matrix $\mathbb{M}(c)$ in
Eq.~\eqref{eigval_eqn} is not Hermitian but rather complex-symmetric. 
How then can we modify the
RSPT to adapt to this situation where
the matrices are complex-symmetric, rather than Hermitian?
We answer this question below.

Because our solution to the complex-symmetric matrix
eigenvalue perturbation problem
is a modification of the popular solution to the  
Hermitian matrix eigenvalue perturbation problem 
(typically encountered in quantum mechanics),
we have chosen to use 
Dirac's bra-ket notation, but with a slight modification. 
$\ket{a}$ as usual, represents a column
vector but $\bra{a}$ will stand for the transpose of $\ket{a}$, 
rather than its conjugate transpose. Hence, 
\begin{equation}
\langle a \mid b\rangle \equiv a^{T} b, ~~~~~~\langle a|\mathbb{M}| b\rangle \equiv a^{T} \mathbb{M} b    .
\end{equation}
Now, the unperturbed and the perturbed eigenvalue equations are
\begin{align}
H_0\left|n_0\right\rangle   & =   E^0\left|n_0\right\rangle    ,  \\
\left(  H_{0} + V \right)|n\rangle  & =  ( E^{0}  +  \Delta)|n\rangle  ,
\end{align}
respectively. Here $H_0$ and $E^0$ are the unperturbed complex-symmetric 
matrix, and the associated unperturbed eigenvalue, 
respectively.
We have used  $\ket{n_0}$ to denote the unperturbed eigenvector.
 $V$, $\Delta$ and $\ket{n}$ are the complex-symmetric 
perturbation matrix, the perturbation in the eigenvalue, and
 the perturbed eigenvector, respectively.
The above equation gives us
\begin{equation}   \label{eq100}
\left\langle n_{0}\left|\left(E^{0}-H_{0}\right)\right| n\right\rangle=\left\langle n_{0}|(V-\Delta)| n\right\rangle  .   
\end{equation}
Since $H_0$ is a complex-symmetric matrix, it can be easily shown for any 
general vectors $\ket{a}$ and $\ket{b}$ with complex entries, that
$\left\langle a\left|H_{0}\right| b\right\rangle=\left\langle b\left|H_{0}\right| a\right\rangle$. 
This, along with $\braket{a|b} = \braket{b|a}$
makes it possible to write Eq.~\eqref{eq100} as
\begin{align}
\left\langle n_{0}|(V-\Delta)| n\right\rangle    = &  \left\langle n_{0}\left|E^{0}\right| n\right\rangle-\left\langle n_{0}\left|H_{0}\right| n\right\rangle \nonumber \\
 =  & \left\langle n_{0}\left|E^{0}\right| n\right\rangle-\left\langle n\left|H_{0}\right| n_{0}\right\rangle      \nonumber  \\
 =  &  E^{0}\left(\left\langle n_{0} \mid n\right\rangle-\left\langle n \mid n_{0}\right\rangle\right)=0  .
\end{align}
This finally yields
\begin{equation}
\begin{aligned}
&\left\langle n_{0}|(V-\Delta)| n\right\rangle=0 \\
&\Delta=\frac{\left\langle n_{0}|V| n\right\rangle}{\left\langle n_{0} \mid n\right\rangle}=\frac{\left\langle n_{0}|V| n_{0}\right\rangle}{\left\langle n_{0} \mid n_{0}\right\rangle} .
\end{aligned}
\end{equation}
To arrive at the last result, we 
replaced $\ket{n}$ with $\ket{n_0} $. This is justified
 because it makes a difference at a higher order in 
 perturbation scheme than what we are concerned here.

\section{Software package}     \label{python_package}

We have implemented the improvements suggested in Sec.~\ref{our-method},
based on the first derivative information in a
Git-forked version of the original \textsc{Python} package \texttt{qnm} 
that was introduced with Ref.~\cite{Stein:2019mop}.
This forked repository can be found at Ref.~\cite{tanay_python}. 
At a certain fixed BH spin $a$, preliminary 
investigation suggests that with these improvements, 
the root-finding  procedure is faster by a factor of 1.5 when compared with the original  \texttt{qnm}
package of Ref.~\cite{Stein:2019mop}. Since this is done only at a fixed $a$, the 
analytical derivatives $d \omega/da$ and $dA/da$ do not matter. This speed boost  
could be due to supplying the analytical derivative $d \cf/d \omega$ to the Newton-Raphson routine,
as discussed in Sec.~\ref{our-method}. The reason we do not make comparisons while changing the 
BH spin $a$ is that the original \texttt{qnm} package uses numerical second derivative (in
the form of quadratic fits and extrapolation) to predict $\omega_{\text{guess}}$ and 
$A_{\text{guess}}$ at various $a$'s, whereas the forked version of the package
makes this prediction on the basis of only the first (analytical) derivatives. 
We plan to make these comparisons once the analytical second derivative information
is also included in the package in the future.

\section{Summary}      \label{conc}

In this paper, we have laid the theoretical groundwork to make the
spectral variant of Leaver's method of finding the QNM frequencies more robust.
This is done by introducing ways to compute the necessary derivatives analytically
rather than numerically. We have confined ourselves to first derivatives only, 
leaving the work on second derivatives for future work. Our work should be 
particularly useful when the QNM frequency trajectory (as the BH spin $a$ is varied)
 in the complex plane has large curvature. We have incorporated the first
 derivative information in the \textsc{Python} package \texttt{qnm}.

We saw in Sec.~\ref{standard_sec} that inclusion of 
the second derivative information can let us take large step size in the BH spin $a$.
This, combined with the fact that it is the second derivative (rather 
than the first derivative) that lets us decide the right step size to take when using 
the adaptive step size approach,  makes the prospect of extending our present work
to include  second derivatives, the most natural future course of action.
Thereafter, one can include the analytical second derivative information too
in the \texttt{qnm} \textsc{Python} package that was
introduced with Ref.~\cite{Stein:2019mop}.
Comparisons can then be made to assess the improvements brought about by 
computing derivatives analytically rather than numerically. These 
can be in regard to the speed of finding the QNM frequencies, the size of the 
step in the BH spin $a$ that we can safely take, the robustness of the routine
(can it find the QNM frequencies that the present routines can't), among others.
There is not much point in pushing our derivative calculations to very high 
orders. This is because after a certain point, due to the high order of the analytical derivative being 
computed, the computational overhead will start to outweigh the gains it may bring.
Finally, there is also hope to adapt our work to find QNM frequencies
in some modified theories of gravity, provided in those theories, the QNM 
frequencies are found using a method that is  a variant of Leaver's method.

%\acknowledgments

%The work of L.C.S. was partially supported by NSF
%CAREER Award PHY-2047382.

%  \appendix   

\bibliography{bib_file}

%apsrev4-2.bst 2019-01-14 (MD) hand-edited version of apsrev4-1.bst
%Control: key (0)
%Control: author (8) initials jnrlst
%Control: editor formatted (1) identically to author
%Control: production of article title (0) allowed
%Control: page (0) single
%Control: year (1) truncated
%Control: production of eprint (0) enabled
\begin{thebibliography}{10}%
\makeatletter
\providecommand \@ifxundefined [1]{%
 \@ifx{#1\undefined}
}%
\providecommand \@ifnum [1]{%
 \ifnum #1\expandafter \@firstoftwo
 \else \expandafter \@secondoftwo
 \fi
}%
\providecommand \@ifx [1]{%
 \ifx #1\expandafter \@firstoftwo
 \else \expandafter \@secondoftwo
 \fi
}%
\providecommand \natexlab [1]{#1}%
\providecommand \enquote  [1]{``#1''}%
\providecommand \bibnamefont  [1]{#1}%
\providecommand \bibfnamefont [1]{#1}%
\providecommand \citenamefont [1]{#1}%
\providecommand \href@noop [0]{\@secondoftwo}%
\providecommand \href [0]{\begingroup \@sanitize@url \@href}%
\providecommand \@href[1]{\@@startlink{#1}\@@href}%
\providecommand \@@href[1]{\endgroup#1\@@endlink}%
\providecommand \@sanitize@url [0]{\catcode `\\12\catcode `\$12\catcode
  `\&12\catcode `\#12\catcode `\^12\catcode `\_12\catcode `\%12\relax}%
\providecommand \@@startlink[1]{}%
\providecommand \@@endlink[0]{}%
\providecommand \url  [0]{\begingroup\@sanitize@url \@url }%
\providecommand \@url [1]{\endgroup\@href {#1}{\urlprefix }}%
\providecommand \urlprefix  [0]{URL }%
\providecommand \Eprint [0]{\href }%
\providecommand \doibase [0]{https://doi.org/}%
\providecommand \selectlanguage [0]{\@gobble}%
\providecommand \bibinfo  [0]{\@secondoftwo}%
\providecommand \bibfield  [0]{\@secondoftwo}%
\providecommand \translation [1]{[#1]}%
\providecommand \BibitemOpen [0]{}%
\providecommand \bibitemStop [0]{}%
\providecommand \bibitemNoStop [0]{.\EOS\space}%
\providecommand \EOS [0]{\spacefactor3000\relax}%
\providecommand \BibitemShut  [1]{\csname bibitem#1\endcsname}%
\let\auto@bib@innerbib\@empty
%</preamble>
\bibitem [{\citenamefont {Berti}\ \emph {et~al.}(2009)\citenamefont {Berti},
  \citenamefont {Cardoso},\ and\ \citenamefont {Starinets}}]{Berti:2009kk}%
  \BibitemOpen
  \bibfield  {author} {\bibinfo {author} {\bibfnamefont {E.}~\bibnamefont
  {Berti}}, \bibinfo {author} {\bibfnamefont {V.}~\bibnamefont {Cardoso}},\
  and\ \bibinfo {author} {\bibfnamefont {A.~O.}\ \bibnamefont {Starinets}},\
  }\bibfield  {title} {\bibinfo {title} {{Quasinormal modes of black holes and
  black branes}},\ }\href {https://doi.org/10.1088/0264-9381/26/16/163001}
  {\bibfield  {journal} {\bibinfo  {journal} {Class. Quant. Grav.}\ }\textbf
  {\bibinfo {volume} {26}},\ \bibinfo {pages} {163001} (\bibinfo {year}
  {2009})},\ \Eprint {https://arxiv.org/abs/0905.2975} {arXiv:0905.2975
  [gr-qc]} \BibitemShut {NoStop}%
\bibitem [{\citenamefont {Nollert}(1999)}]{Nollert_1999}%
  \BibitemOpen
  \bibfield  {author} {\bibinfo {author} {\bibfnamefont {H.-P.}\ \bibnamefont
  {Nollert}},\ }\bibfield  {title} {\bibinfo {title} {Quasinormal modes: the
  characteristic {\textasciigrave}sound{\textquotesingle} of black holes and
  neutron stars},\ }\href {https://doi.org/10.1088/0264-9381/16/12/201}
  {\bibfield  {journal} {\bibinfo  {journal} {Classical and Quantum Gravity}\
  }\textbf {\bibinfo {volume} {16}},\ \bibinfo {pages} {R159} (\bibinfo {year}
  {1999})}\BibitemShut {NoStop}%
\bibitem [{\citenamefont {Leaver}(1985)}]{Leaver:1985ax}%
  \BibitemOpen
  \bibfield  {author} {\bibinfo {author} {\bibfnamefont {E.~W.}\ \bibnamefont
  {Leaver}},\ }\bibfield  {title} {\bibinfo {title} {{An Analytic
  representation for the quasi normal modes of Kerr black holes}},\ }\href
  {https://doi.org/10.1098/rspa.1985.0119} {\bibfield  {journal} {\bibinfo
  {journal} {Proc. Roy. Soc. Lond. A}\ }\textbf {\bibinfo {volume} {402}},\
  \bibinfo {pages} {285} (\bibinfo {year} {1985})}\BibitemShut {NoStop}%
\bibitem [{\citenamefont {{Cook}}\ and\ \citenamefont
  {{Zalutskiy}}(2014)}]{CZ}%
  \BibitemOpen
  \bibfield  {author} {\bibinfo {author} {\bibfnamefont {G.~B.}\ \bibnamefont
  {{Cook}}}\ and\ \bibinfo {author} {\bibfnamefont {M.}~\bibnamefont
  {{Zalutskiy}}},\ }\bibfield  {title} {\bibinfo {title} {{Gravitational
  perturbations of the Kerr geometry: High-accuracy study}},\ }\href
  {https://doi.org/10.1103/PhysRevD.90.124021} {\bibfield  {journal} {\bibinfo
  {journal} {\prd}\ }\textbf {\bibinfo {volume} {90}},\ \bibinfo {eid} {124021}
  (\bibinfo {year} {2014})},\ \Eprint {https://arxiv.org/abs/1410.7698}
  {arXiv:1410.7698 [gr-qc]} \BibitemShut {NoStop}%
\bibitem [{\citenamefont {Hughes}(2000)}]{Hughes:1999bq}%
  \BibitemOpen
  \bibfield  {author} {\bibinfo {author} {\bibfnamefont {S.~A.}\ \bibnamefont
  {Hughes}},\ }\bibfield  {title} {\bibinfo {title} {{The Evolution of
  circular, nonequatorial orbits of Kerr black holes due to gravitational wave
  emission}},\ }\href {https://doi.org/10.1103/PhysRevD.65.069902} {\bibfield
  {journal} {\bibinfo  {journal} {Phys. Rev. D}\ }\textbf {\bibinfo {volume}
  {61}},\ \bibinfo {pages} {084004} (\bibinfo {year} {2000})},\ \bibinfo {note}
  {[Erratum: Phys.Rev.D 63, 049902 (2001), Erratum: Phys.Rev.D 65, 069902
  (2002), Erratum: Phys.Rev.D 67, 089901 (2003), Erratum: Phys.Rev.D 78, 109902
  (2008), Erratum: Phys.Rev.D 90, 109904 (2014)]},\ \Eprint
  {https://arxiv.org/abs/gr-qc/9910091} {arXiv:gr-qc/9910091} \BibitemShut
  {NoStop}%
\bibitem [{\citenamefont {Stein}(2019)}]{Stein:2019mop}%
  \BibitemOpen
  \bibfield  {author} {\bibinfo {author} {\bibfnamefont {L.~C.}\ \bibnamefont
  {Stein}},\ }\bibfield  {title} {\bibinfo {title} {{qnm: A Python package for
  calculating Kerr quasinormal modes, separation constants, and
  spherical-spheroidal mixing coefficients}},\ }\href
  {https://doi.org/10.21105/joss.01683} {\bibfield  {journal} {\bibinfo
  {journal} {J. Open Source Softw.}\ }\textbf {\bibinfo {volume} {4}},\
  \bibinfo {pages} {1683} (\bibinfo {year} {2019})},\ \Eprint
  {https://arxiv.org/abs/1908.10377} {arXiv:1908.10377 [gr-qc]} \BibitemShut
  {NoStop}%
%%CITATION = ARXIV:1908.10377;%%
\bibitem [{\citenamefont {Press}\ \emph {et~al.}(2007)\citenamefont {Press},
  \citenamefont {Teukolsky}, \citenamefont {Vetterling},\ and\ \citenamefont
  {Flannery}}]{press2007numerical}%
  \BibitemOpen
  \bibfield  {author} {\bibinfo {author} {\bibfnamefont {W.}~\bibnamefont
  {Press}}, \bibinfo {author} {\bibfnamefont {S.}~\bibnamefont {Teukolsky}},
  \bibinfo {author} {\bibfnamefont {W.}~\bibnamefont {Vetterling}},\ and\
  \bibinfo {author} {\bibfnamefont {B.}~\bibnamefont {Flannery}},\ }\href
  {https://books.google.com/books?id=1aAOdzK3FegC} {\emph {\bibinfo {title}
  {Numerical Recipes 3rd Edition: The Art of Scientific Computing}}}\ (\bibinfo
   {publisher} {Cambridge University Press},\ \bibinfo {year}
  {2007})\BibitemShut {NoStop}%
\bibitem [{\citenamefont {Press}\ and\ \citenamefont
  {Teukolsky}(1988)}]{aip_lentz}%
  \BibitemOpen
  \bibfield  {author} {\bibinfo {author} {\bibfnamefont {W.~H.}\ \bibnamefont
  {Press}}\ and\ \bibinfo {author} {\bibfnamefont {S.~A.}\ \bibnamefont
  {Teukolsky}},\ }\bibfield  {title} {\bibinfo {title} {Evaluating continued
  fractions and computing exponential integrals},\ }\href
  {https://doi.org/10.1063/1.4822777} {\bibfield  {journal} {\bibinfo
  {journal} {Computers in Physics}\ }\textbf {\bibinfo {volume} {2}},\ \bibinfo
  {pages} {88} (\bibinfo {year} {1988})},\ \Eprint
  {https://arxiv.org/abs/https://aip.scitation.org/doi/pdf/10.1063/1.4822777}
  {https://aip.scitation.org/doi/pdf/10.1063/1.4822777} \BibitemShut {NoStop}%
\bibitem [{\citenamefont {Griffiths}\ and\ \citenamefont
  {Schroeter}(2018)}]{griffiths}%
  \BibitemOpen
  \bibfield  {author} {\bibinfo {author} {\bibfnamefont {D.}~\bibnamefont
  {Griffiths}}\ and\ \bibinfo {author} {\bibfnamefont {D.}~\bibnamefont
  {Schroeter}},\ }\href {https://books.google.com/books?id=82FjDwAAQBAJ} {\emph
  {\bibinfo {title} {Introduction to Quantum Mechanics}}}\ (\bibinfo
  {publisher} {Cambridge University Press},\ \bibinfo {year}
  {2018})\BibitemShut {NoStop}%
\bibitem [{tan()}]{tanay_python}%
  \BibitemOpen
  \href@noop {} {}\bibinfo {howpublished}
  {\url{https://github.com/sashwattanay/qnm}}\BibitemShut {NoStop}%
\end{thebibliography}%

\end{document}